\documentclass[fleqn,10pt]{wlscirep}
\usepackage[utf8]{inputenc}
\usepackage[T1]{fontenc}
\usepackage{amsthm}
\newtheorem{definition}{Definition}
\usepackage[authoryear]{natbib}

\usepackage{dsfont}

\title{Extended Scale-Free Networks}

\author[1,*]{Arthur Charpentier}
\author[2]{Emmanuel Flachaire}
\affil[1]{Université du Québec à Montréal (UQAM), Canada}
\affil[2]{Aix-Marseille University (AMSE), France}

\affil[*]{corresponding author: \textcolor{blue}{ charpentier.arthur@uqam.ca}}


\begin{abstract}
Recently, \cite{Broido} mentioned that (strict) Scale-Free networks were rare, in real life. This might be related to the statement of \cite{Stumpf}, that sub-networks of scale-free networks are not scale-free. In the later, those sub-networks are asymptotically scale-free, but one should not forget about second-order deviation (possibly also third order actually). In this article, we introduce a concept of extended scale-free network, inspired by the extended Pareto distribution, that actually is maybe more realistic to describe real network than the strict scale free property. This property is consistent with \cite{Stumpf} : sub-network of scale-free larger networks are not strictly scale-free, but extended scale-free.
\end{abstract}
\begin{document}

\flushbottom
\maketitle
%
%
\thispagestyle{empty}

\section{From Strict Scale-Free to Scale-Free Types}

scale-free network is a network whose degree distribution follows a power law, at least asymptotically. When studying internet networks, \cite{Barabasi} observed that some nodes, that they called {\em hubs}, had much more connections than others, and that the distribution of the number of links connecting to a node was a power-law. They coined the term {\em scale-free network} to describe that class of networks, when degrees have a power-law distribution. \cite{Clauset} studied real networks, and found some exhibiting that property. Nevertheless, recently, \cite{Broido} claimed that (strict) scale-free network are actually rare. More specifically, inspired by \cite{Alderson}, they define various notions of {\em weak} or {\em strong} scale-free networks. If their taxonomy of scale-free network, in interesting we will consider here only the concept of {\em strict} scale-free if the degree distribution above a given cutoff $k_{\min}$ is a power law (as in \cite{Barabasi}).

As discussed in \cite{Voitalov}, the scale-free property is closely related to the Pareto distribution, used in extreme value theory (and is the continuous version of the discrete power-law used for degrees). Nevertheless, as we will recall, this (strict) Pareto appears usually only above a very high threshold, and distributions are only Pareto {\em type}. Recently, \cite{Beirlant} suggested to take into account the second order approximation: the first order has a power law, and so is the second order, with smaller tail index. This is the extended Pareto distribution. In this article, we will explore an ``{\em extended scale-free}'' property, and study its impact on networks. And as we will see, this property is close to one described in \cite{Stumpf} when studying sub-networks of larger ones.

\subsection{Strict Scale-Free}

Consider a network $(\mathcal{V},\mathcal{E})$, and let $n$ denote the number of nodes. Following \cite{Barabasi}, the network is said to be scale-free if
$$
p_{k_{\min}}(k)=\mathbb{P}[D=k]=Ck^{-\alpha},~\alpha>1,\text{ for } k\geq k_{\min}\geq 1
$$
where $\alpha$ is the scaling exponent, and $C$ is the normalization constant. Inference is performed using a degree sequence $d_1,\cdots,d_n$. Equivalently, the log-log plot should be linear (and the absolute value of the slope is the scaling exponent)
$$
\log p_{k_{\min}}(k) = \log c -\alpha \log k.
$$
This linear property for the logarithm of the frequency is the one usually used in network studies. Note that $p_{1}$ is also called Zipf's distribution.

\begin{definition}
The discrete Zipf's distribution is defined as
$$
p_{d-\text{Z},\alpha}(k)=\frac{1}{\zeta(\alpha)}k^{-\alpha}\text{ for }k\in\mathbb{N}_+=\lbrace1,2,\ldots\rbrace
$$
where $\zeta()$ is Rieman's function, $\displaystyle{\zeta(\alpha)=\sum_{k=1}^\infty k^{-\alpha}}$, defined for $\alpha>1$.
\end{definition}

In this article, we exclude nodes that have $0$ connection. In many applications, some nodes are just disconnected from the others, having a null degree. In social networks, those individuals might be simple observers, and do not interact with others. This will also appear when considering sub-networks. Hence, when dealing with sub-networks in the last section of this article, we will remove nodes that are not connected to anyone.

\subsection{Discrete PD with Cumulative Probabilities}

Another popular approach is to consider cumulative probabilities, instead of frequencies. An interesting feature is that the cumulative probability of a power law probability distribution is also power law, but with an exponent $\alpha-1$.
More specifically, let $\overline{F}(k)=\mathbb{P}[D>k]=1-\mathbb{P}[D\leq k]$ denote the complementary cumulative distribution function. If we consider a continuous version of $p(x)=Cx^{-\alpha}$, we obtain
$$
\overline{F}(x)=\int_x^\infty p(t)dt=C\int_x^\infty t^{-\alpha}dt=\frac{Cx^{\alpha-1}}{\alpha-1}=\gamma x^{\alpha-1}
$$
which is also a power function. And it is actually possible to derive a discrete probability function from a power $\overline{F}$ function, or more such as the (standard) Pareto function (see \cite{Arnold} or chapter 20 in \cite{JohnsonKotz})
$$
\overline{F}_{\text{PD},\xi}(x)=x^{-1/\xi}\text{ for } x\geq 1, 
$$
with $\xi>0$.

\begin{definition}
The discrete strict Pareto distribution is defined as
$$
p_{d-\text{PD},\xi}(k)=k^{-1/\xi}-(k+1)^{-1/\xi}\text{ for }k\in\mathbb{N}_+
$$
defined for $\xi\in\mathbb{R}_+$.
\end{definition}

For discretized version of continuous ones, we will use tail index $\xi$, having in mind the fact that $\alpha$ is of order $1+1/\xi$. The popular case $\alpha\in(2,3)$ means that $\xi\in(1/2,1)$, with $\alpha$ all the more small that $\xi$ is large. Observe that if the degrees have a discrete strict Pareto distribution, their expected value is
$$
<D_{d-\text{P}}>=\sum_{k=1}^\infty kp_{d-\text{PD},\xi}(k)=\sum_{k=1}^\infty k^{-1/\xi}=\zeta(1/\xi)
$$
which is different from $(1-\xi)^{-1}$ obtained with a continuous Pareto distribution. 

\subsection{Discrete GPD and Second Law of Extremes}

Pareto distributions are very popular since the second law of extremes (see \cite{Pickands} and \cite{Balkema}) which states that if $X$ is a random variable such that there exists a function $a(u)$ such that
$$
a(u)^{-1}(X-u)|X>u \rightarrow Z, \text{ as } u\rightarrow\infty
$$
(in the {\em weak convergence} sense) for some non-degenerate $Z$ on $\mathbb{R}_+$, then $Z$ follows a Generalized Pareto (GPD) with complementary cumulative distribution function
$$
\overline{F}_{\text{GPD},\sigma,\xi}(x)=\mathbb{P}[Z>x]=\left(1+\xi\frac{x}{\sigma}\right)^{-1/\xi},\text{ for } x\geq 0. 
$$
As a consequence, for large $u$, we have usually the approximation for $\mathbb{P}[X-u>x|X>u]$
\begin{eqnarray*}
\mathbb{P}[a(u)\big(a(u)^{-1}(X-u)\big)>x|X>u] 
\sim\mathbb{P}[a(u)Z>x]= \mathbb{P}[Z>a(u)^{-1}x]
= \overline{F}_{\text{GPD},\sigma a(u),\xi}(x)
\end{eqnarray*}
as suggested in \cite{Davison}. We will then write $X\in \text{MDA}_{\xi}$ - for {\em Max-Domain of Attraction}.
Nevertheless, as proved in \cite{Anderson} and \cite{Shimura}, this approximation might not be valid if $X$ as a discrete support. An important additional property if to have a long-tailed distribution for $\overline{F}$  in the sens that 
$$
\frac{\overline{F}(u+1)}{\overline{F}(u)}\rightarrow 1\text{ as }u\rightarrow \infty
$$
As proved in \cite{Shimura}, a discrete random variable $X\in \text{MDA}_{\xi}$ if and only if (i) $\overline{p}$ is long-tailed and (ii) $X=\lceil X^\star\rceil$ where $X^\star\in \text{MDA}_{\xi}$. And in that case, 
$$
p_{u}(u+k)=\mathbb{P}[X-u=k|X>u]\approx p_{d-\text{GPD},a(u)\sigma,\xi}(k),\text{ for }k\in\mathbb{N}
$$
Thus, following \cite{Krishna}, define :
\begin{definition}
The discrete generalized Pareto distribution is defined as
$$
p_{d-\text{GPD},\sigma,\xi}(k)=\overline{F}_{\text{GPD},\sigma,\xi}(k-1)-\overline{F}_{\text{GPD},\sigma,\xi}(k)=\left(1+\xi\frac{k-1}{\sigma}\right)^{-1/\xi}-\left(1+\xi\frac{k}{\sigma}\right)^{-1/\xi}\text{ for }k\in\mathbb{N}_+
$$
defined for $\xi\in\mathbb{R}_+$.
\end{definition}

A (continuous) GPD random variable can be expressed as a mixture of exponential variables, that is an exponential random variable with a Gamma distributed rate parameter : if $X^\star\sim\mathcal{E}(\Lambda)$ with $\Lambda\sim\mathcal{G}(\alpha,\beta)$, then $X^\star$ has a GPD distribution, with tail index $\xi=1/\alpha$. Interestingly, as proved in \cite{Buddana}, a similar property holds for the discrete $d$-GPD, which is a mixture of geometric variables (with also Gamma heterogeneity). 

\begin{figure}
    \centering
    \includegraphics[width=.8\textwidth]{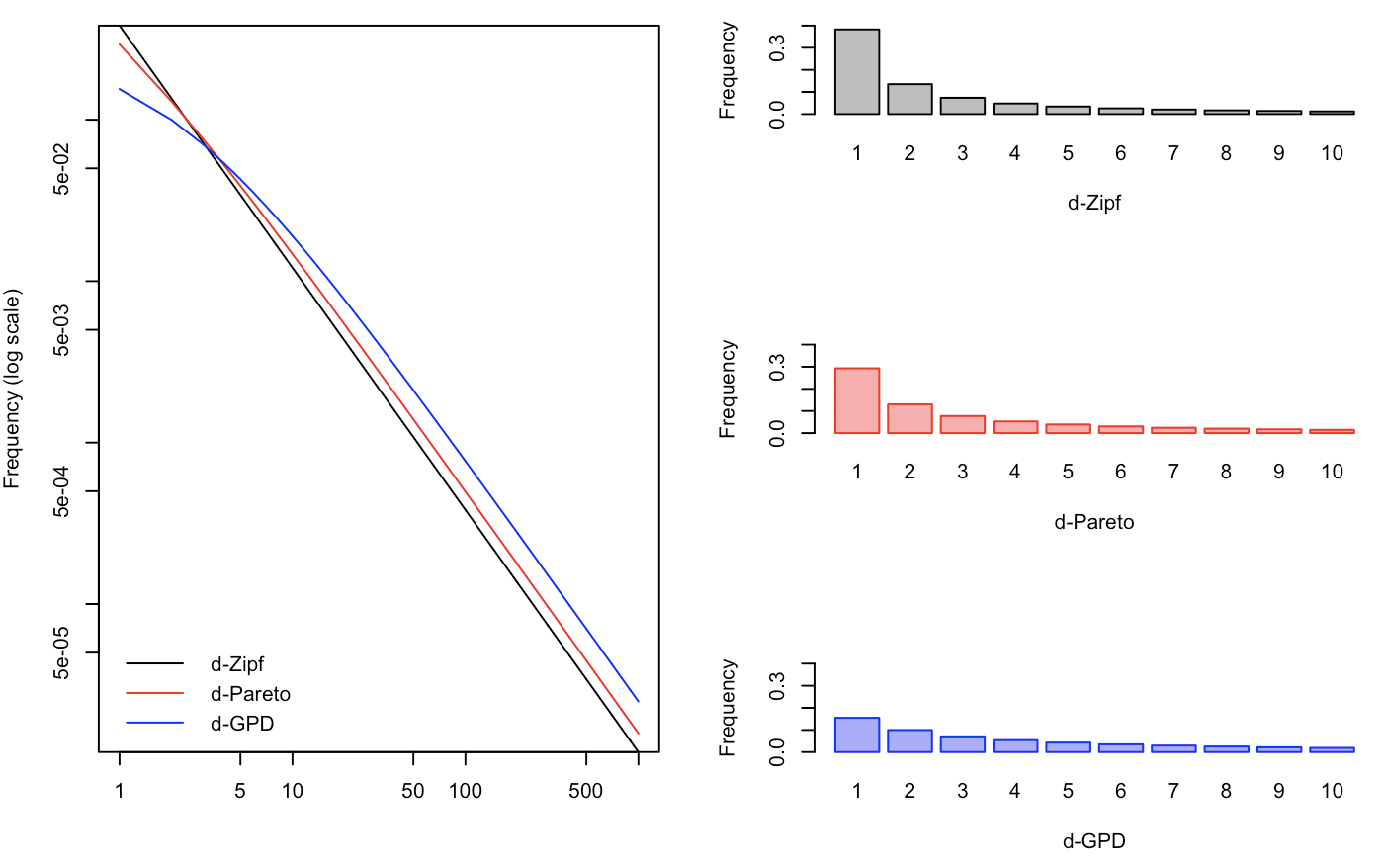}
    \caption{$p_{d-\text{Z}}$, $p_{d-\text{P}}$ and $p_{d-\text{GPD}}$, on a log-log scale on the left, and for the first 10 values on the right, with the same index $\xi$.}
    \label{fig:p:1}
\end{figure}

The three probability distributions ($p_{d-\text{Z}}$, $p_{d-\text{P}}$ and $p_{d-\text{GPD}}$), for similar tail exponent $\xi$, can be visualized on Figure \ref{fig:p:1} (for the Zipf, $\alpha=1+1/\xi$).

\subsection{Regular Variation and Power-Law Type}

This power law property is deeply related to the concept of {\em scale-free} distribution :  scale-free means that the distribution is the same whatever scale we consider. Hence, for any $\lambda$, $f(\lambda k)$ is proportional to $f(k)$, or  $f(\lambda k)=h_\lambda f(x)$ (we must consider here the continuous version $f$ and not $p$ since, unless $\lambda\in\mathbb{N}$, $\lambda k$ is not always an integer). Thus, since $h_\lambda=f(\lambda)/f(1)$, we get that $f(\lambda x)=f(\lambda)\cdot f(x)$, which is the multiplicative version of Cauchy's functional equation (also called Hamel-Cauchy), with unique solution $f(x)=x^{-\alpha}$ (up to a multiplicative constant). Hence, scale-free means that $f(\lambda x)/f(x)$ is constant (and in that case, $f(x)=C x^{-\alpha}$). A natural extension is to assume that $f(\lambda x)/f(x)$ is asymptotically constant.

A function $g$ is said to be regularly varying (at infinity) if $g(tx)/g(x)$ tends to $t^\theta$, for some $\theta\in\mathbb{R}$, when $t$ goes to infinity. If $\theta=0$, then $g$ is said to be slowly varying, to derive an extended version of the power law.

\begin{definition}
A continuous variable $X^\star$ is said to be Pareto-type distributed, with tail exponent $\xi$ if $\mathbb{P}[X^\star>x]=x^{-1/\xi}\ell(x)$ for some slowly varying function $\ell$.
\end{definition}

In section \ref{section:ESF}, the idea will be to consider a simple parametric expresion for function $\ell$, that will decay to a constant at some power speed.

\subsection{Probability-Generating Function of Scale-Free Distribution}

An alternative way to describe the distribution is not to use $p$, but its probability-generating function (PGF), $G(s)$, defined as
$$
G(s)=\sum_{k=0}^\infty p(k)s^k,
$$
for instance, with a Poisson variable, ${\displaystyle G(z)=e^{\lambda (z-1)}} $, while with a power law, or a scale free distribution
$$
G(s)=\frac{1}{\zeta(\alpha)}\sum_{i=1}^\infty i^{-\alpha}s^i=\frac{\text{Li}_\alpha(s)}{\zeta(\alpha)}
$$
where $\text{Li}_\alpha$ is Jonquière's polylogarithm function (see \cite{Abramowitz}). We will use that alternative representation when focusing on sub-networks.

\section{Extended Scale-Free}\label{section:ESF}

The Extended Pareto Distribution (EPD) was introduced in \cite{Beirlant}, and there are many way to derive that distribution, most of them being equivalent.

\subsection{Mixture of Scale-Free}

\cite{Hall} suggested to write a Pareto type distribution $\overline{F}(x)=x^{-1/\xi}\ell(x)$ as  $\overline{F}(x)=x^{-1/\xi}\cdot C(1-\delta(x))$. Here, $\ell$ is not only slowly varying, but also $\ell(x)$ tends to $C$ when $x$ goes to infinity (at some power rate). More specifically, assume that $\delta(x)=Dx^\beta+o(x^\beta)$ where $\beta<0$. If $\delta(x)=Dx^\beta$, we can write
$$
\overline{F}(x)=C_1x^{-\gamma_1}+C_2x^{-\gamma_2}
$$
which can be seen as a mixture of two (strict) Pareto distributions.


\subsection{Second-Order Regular Variation}

The first law of extremes (also called Fisher-Tippett theorem) is based on the limiting distribution of maximum $x_{(n)}$ of an i.i.d. sample $\{x_1,\cdots,x_n\}$. More precisely, assume that there exists a function $a(n)$ such that
$$
a(n)^{-1}\big[X_{(n)}-F^{-1}(1-1/n)\big] \rightarrow Z, \text{ as } n\rightarrow\infty
$$
for some non-degenerate $Z$ on $\mathbb{R}_+$, then $Z$ follows a Generalized Extreme Value (GEV) distribution (see \cite{EKM} or \cite{BS}). Let $U$ denote the quantile function $U(x)=F^{-1}(1-1/x)$, then somehow, we might be interested by the limit (if it exists) of $a(n)^{-1}(U(xn)-U(n))$ when $n$ goes to infinity. This is related to the concept of extended regular variation (see \cite{dehaan}) : $g$ is said to be $ERV_\gamma$ if
$$
\lim_{t\rightarrow\infty}\frac{g(tx)-g(t)}{a(t)}=c\frac{x^\gamma-1}{\gamma}
$$
which can be seen as extension of regular variation of index $\gamma$. For instance, the quantile function of a (strict) Pareto distribution with index $\xi$, $U(x)=x^\xi$, and with auxiliary function $a(t)=\xi t^\xi$, $a(n)^{-1}(U(xn)-U(n))=(x^\xi-1)/\xi$ (see \cite{BS}).

Second order regular variation is obtained assuming that there is a function $b$ such that 
$$
\lim_{t\rightarrow\infty}\frac{1}{b(t)}\left[\frac{g(tx)-g(t)}{a(t)}-\frac{x^\gamma-1}{\gamma}\right]
$$
exists, and is denoted $h(x)$. \cite{Stadtmuller} obtained a general expression for $h$, related to some index $\rho$. In a nutshell, following \cite{Drees} and \cite{Cheng}, the limit can be expressed
$$
\lim_{t\rightarrow\infty}\frac{1}{b(t)}\left[\frac{g(tx)-g(t)}{a(t)}-\frac{x^\gamma-1}{\gamma}\right] =\frac{x^{\gamma+\rho}-1}{\gamma+\rho}
$$
with $\rho<0$ (theorem B.3.10 in \cite{ReIns}). 

\subsection{Extended Scale-Free}

For the strict Pareto distribution, we have seen that
$$
\lim_{t\rightarrow\infty}x^{-\gamma}\frac{\overline{F}(tx)}{\overline{F}(t)}=1
$$
But let us consider the following extension (based on the expression of the second-order regular variation)
$$
\lim_{t\rightarrow\infty}x^{-\gamma}\frac{\overline{F}(tx)}{\overline{F}(t)}=1+\frac{x^{\rho}-1}{\rho}, \text{ for some } \rho \leq 0
$$
or, up to some affine transformation, $
\overline{F}(x)=c x^{-\gamma}[1+x^{\rho}\ell(x)]$. 
Since $(1+u)^b\sim (1+bu)$, define (changing $\gamma$ in $\alpha$, $\rho$ in $\tau$)
$$
 \overline{F}(x)=\mathbb{P}[X>x] ={\left[x \left(1+\delta-\delta x^\tau\right)\right]^{-1/\xi}},\text{ for } x\geq u
$$
where $\tau\leq 0$ and $\delta > \max(-1,1/\tau)$. This is the Extended Pareto Distribution, as define in \cite{Beirlant}.

\begin{definition}
The discrete extended Pareto distribution is defined as
$$
p_{d-\text{EPD},\delta,\tau,\xi}(k)=\overline{F}_{\text{EPD},\delta,\tau,\xi}(k-1)-\overline{F}_{\text{EPD},\delta,\tau,\xi}(k)={\left[(k-1) \left(1+\delta-\delta \left(k-1\right)^\tau\right)\right]^{-1/\xi}}-{\left[k \left(1+\delta-\delta k^\tau\right)\right]^{-1/\xi}}\text{ for }k\in\mathbb{N}_+
$$
\end{definition}

\begin{figure}
    \centering
    \includegraphics[width=.6\textwidth]{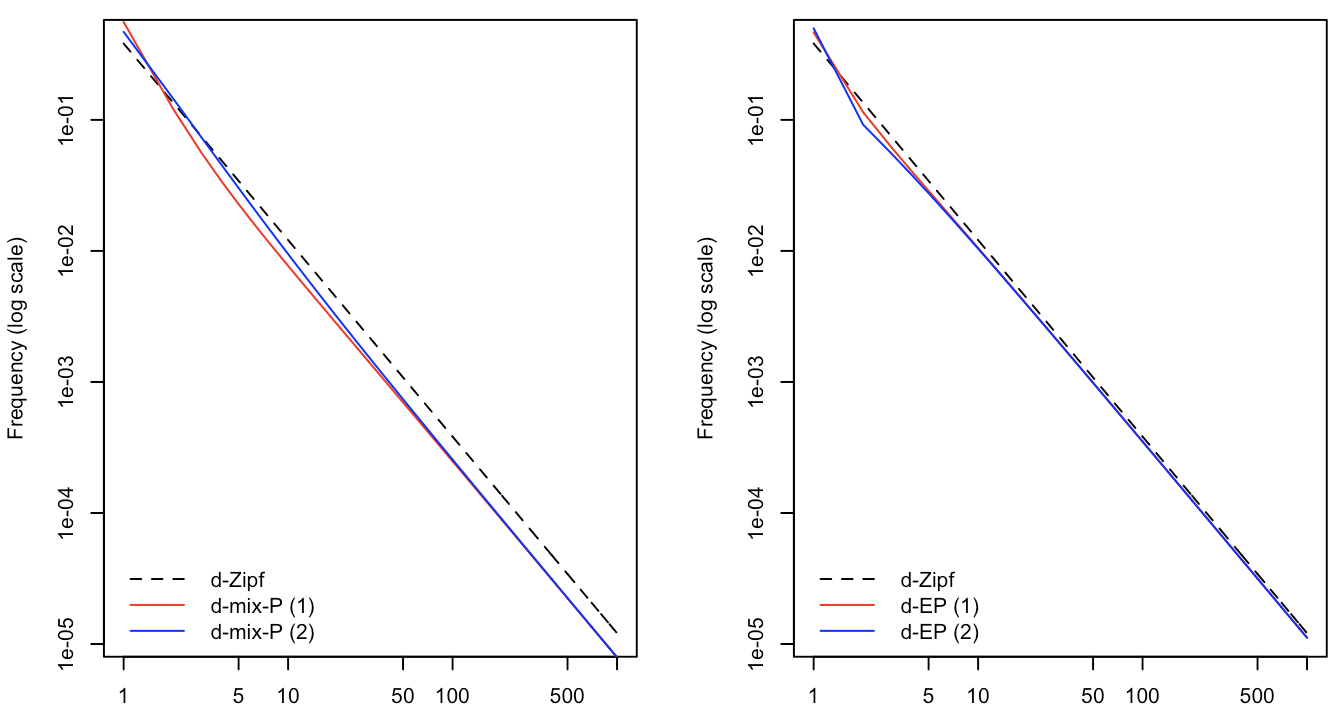}
    \caption{$p_{d-\text{Z}}$ and $p_{d-\text{mix-P}}$ on a log-log scale on the left, and $p_{d-\text{Z}}$ and $p_{d-\text{EP}}$ on the right.}
    \label{fig:p:2}
\end{figure}

\begin{figure}
    \centering
    \includegraphics[width=.6\textwidth]{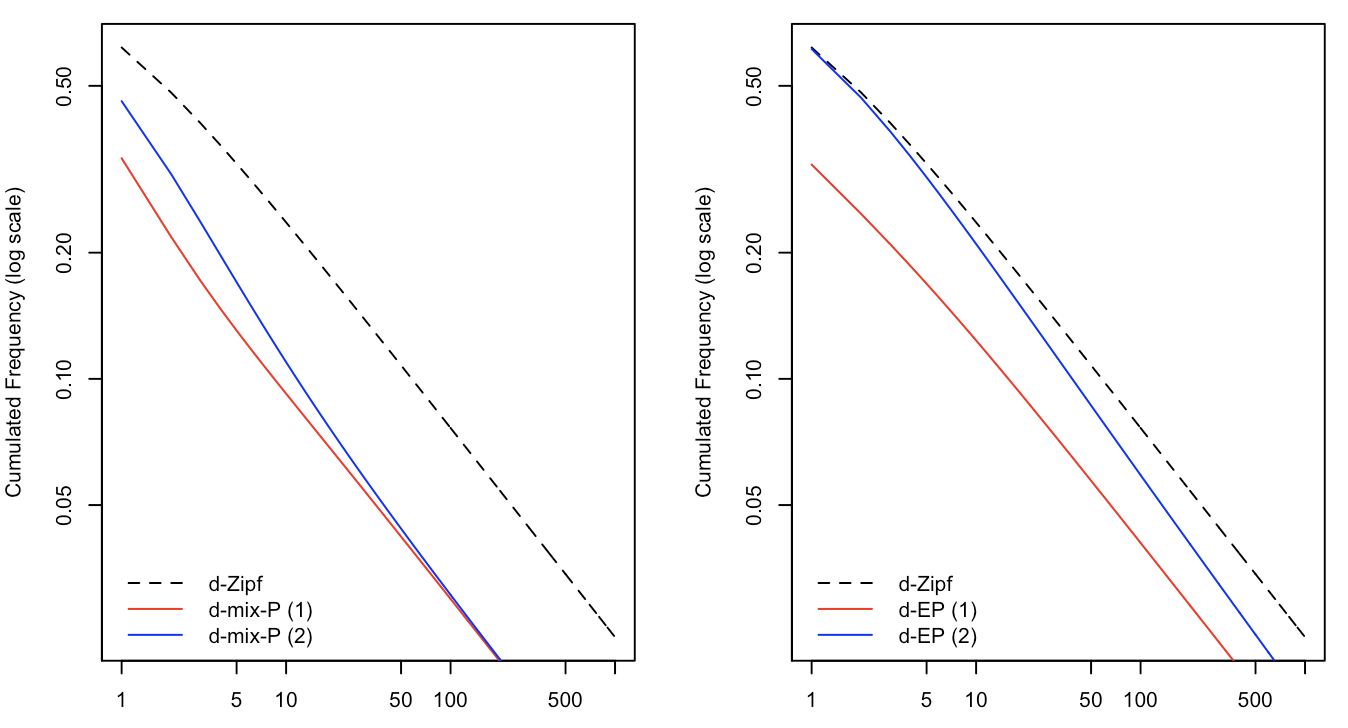}
    \caption{$\overline{F}_{d-\text{Z}}$ and $\overline{F}_{d-\text{mix-P}}$ on a log-log scale on the left, and $\overline{F}_{d-\text{Z}}$ and $\overline{F}_{d-\text{EP}}$ on the right.}
    \label{fig:cdf:2}
\end{figure}

\subsection{Shifted Pareto Distributions}

So far, we defined (discrete) distributions for degrees taking values in $\{1,2,3,\ldots\}$. Quite naturally, one can that $D$ has a Pareto distribution with a shift of $u\in\mathbb{N}_+$ if $D-u$ has a Pareto distribution. For instance, with a strict Pareto distribution, when plotting the complementary cumulative probability function $\overline{F}_{d-\text{PD}}$ on a log-log scale, the function is a (semi)-straight line with slope $-1/\xi$, starting in $(u,2^{-1/\xi})$.

\section{Inference \& Estimation of $\alpha$ or $\xi$}

\subsection{Inference for Continuous Pareto Distributions (Hill Estimator)}

In order to estimate $\alpha$, or $1/\xi$, the power exponent, we can use classical estimators obtained on continuous observations. More specifically, for a strict Pareto sample, use Hill estimate, given a sample $\{x_1,\cdots,x_n\}$, sorted, such that $x_{(1)}\leq x_{(2)}\leq\cdots\leq x_{(n)}$,
$$
\widehat{\xi}=\frac{1}{n}\sum_{i=1}^n\log x_{(i)}-\log x_{(1)}
$$
(see Appendix \ref{App1} for a brief justification) but one can also focus on the $k$ largest values
$$
\widehat{\xi}_k=\frac{1}{k}\sum_{i=n-k+1}^n\log x_{(i)}-\log x_{(n-k)}
$$
This estimator is strongly consistent,  $\widehat{\xi}_k\overset{a.s.}{\rightarrow}\xi$ and (with further assumptions, see \cite{EKM})
$$
\sqrt{k}\big(\widehat{\xi}_k-\alpha\big)\overset{\mathcal{L}}{\rightarrow}\mathcal{N}(0,\xi^2)
$$

\begin{figure}
    \centering
    \includegraphics[width=.8\textwidth]{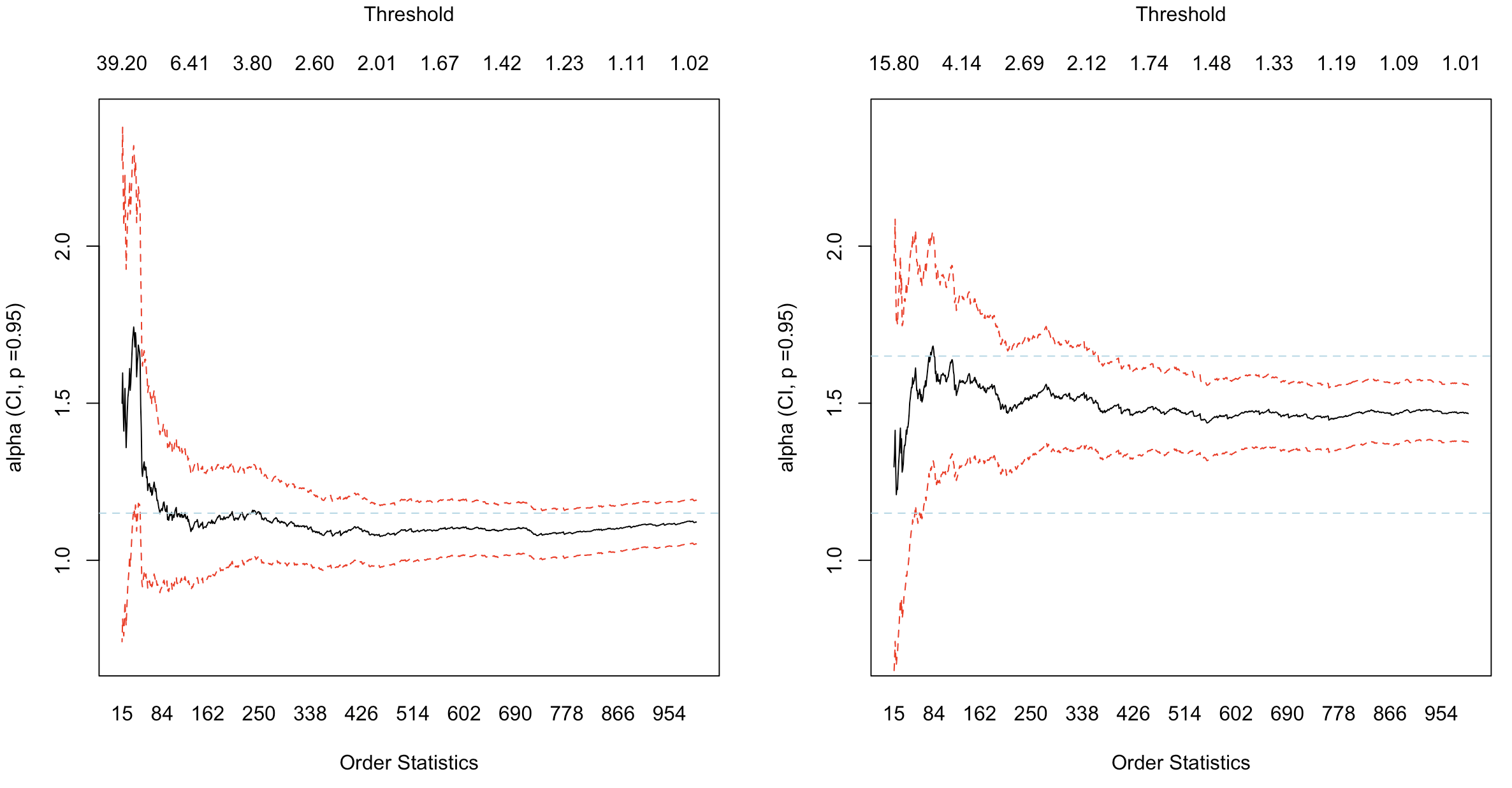}
    \caption{Hill plot, for a (strict) scale free distribution on the left, and an extended scale-free distribution on the right.}
    \label{fig:Hill}
\end{figure}

Nevertheless, this estimator performs badly when the sample is not strictly Pareto distributed, see Figure \ref{fig:Hill}. For the EPD, \cite{ReIns} suggest to use maximum likelihood techniques.

\subsection{Inference for Discrete Pareto Distributions}\label{sec:inf:discrete}

Two techniques are used to estimate parameters (whatever the underlying scale-free distribution considered). The first one is based on the chi-square statistic,
$$
Q(\boldsymbol{\theta})=\sum_{k=1}^{k_{\max}}\frac{(n\cdot p_{d-\star,\boldsymbol{\theta}}(k)-n_k)^2}{n\cdot p_{d-\star,\boldsymbol{\theta}}(k)}
$$
where $n_k$ is the number of nodes with exactly $k$ neighbors. Actually, to get a more robust version, if $n_k$ is too small, we will regroup per classes, to have (at least) 10 nodes per class (see Appendix \ref{App2}). An alternative is to use maximum likelihood techniques (see Appendix \ref{App2}).

\section{Strict and Extended Scale-Free Networks}

Before studying real networks, let us generate networks that are extended scale-free, to see what they look like.

\subsection{Generating a Network from Degree Distribution}

Consider a sequence $d_1,\cdots,d_n$ of non-negative integers such that $d_n\leq\cdots\leq d_1$. From Erd\"os-Gallai theorem, see \cite{Tripathi}, that sequence can be represented as the degree sequence of a finite simple graph on $n$ vertices if and only if $d_1+\cdots+d_n$ is even, and
$$
{\displaystyle \sum _{i=1}^{k}d_{i}\leq k(k-1)+\sum _{i=k+1}^{n}\min(d_{i},k)}
$$
holds for every ${\displaystyle 1\leq k\leq n} $. In this section, we use the methodology described in \cite{Newman} to generate graphs with an Extended Pareto distribution for the degree\footnote{Implemented in the \texttt{graph} library, in {\sffamily R}, see \cite{Gentleman}.}

\subsection{Network Structures}

\begin{figure}
    \centering
    \includegraphics[height=.3\textheight]{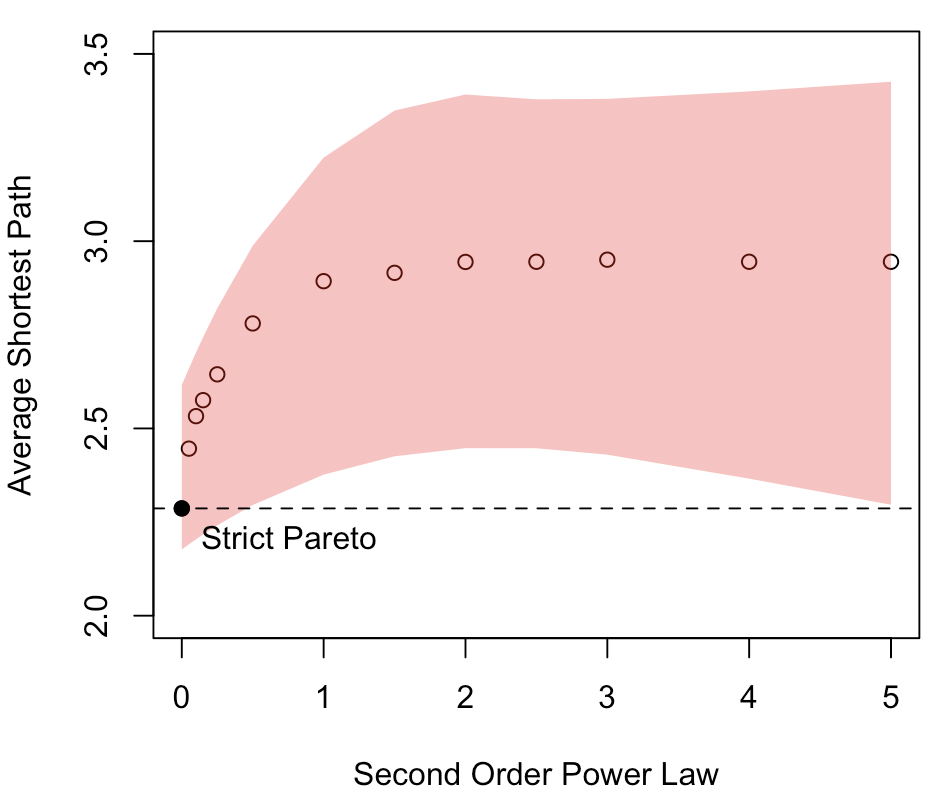}
    \caption{Average shortest path on simulated networks with $n=1,000$ notes, when degrees have an extended scale free distribution $EPD(\gamma=1.15,\kappa,\tau)$ when $-\tau$ varies from $0$ (strict Pareto) to $5$. The shaded area is the $90\%$ confidence band obtained with 500 simulated networks (for each $\tau$).}
    \label{fig:average-shortest-pather}
\end{figure}

On Figure \ref{fig:average-shortest-pather}, we can see the average shortest path for all nodes in the largest connected subgraph. This was obtained by averaging 1,000 simulated networks with $n=1,000$ nodes, with various $\tau$. The larger the absolute value of $\tau$, the longer the average shortest-path distance.

Heuristically, this can be explained since with a strict power law distribution, sub-graphs are connected to each other through (big) hubs, and those network have a small-world property : everyone is close to anyone. With a (strong) second order, there are less very big hubs, and more smaller one : the distance w
to anyone tends to be, on average, longer.

\begin{figure}
    \centering
    \includegraphics[height=.3\textheight]{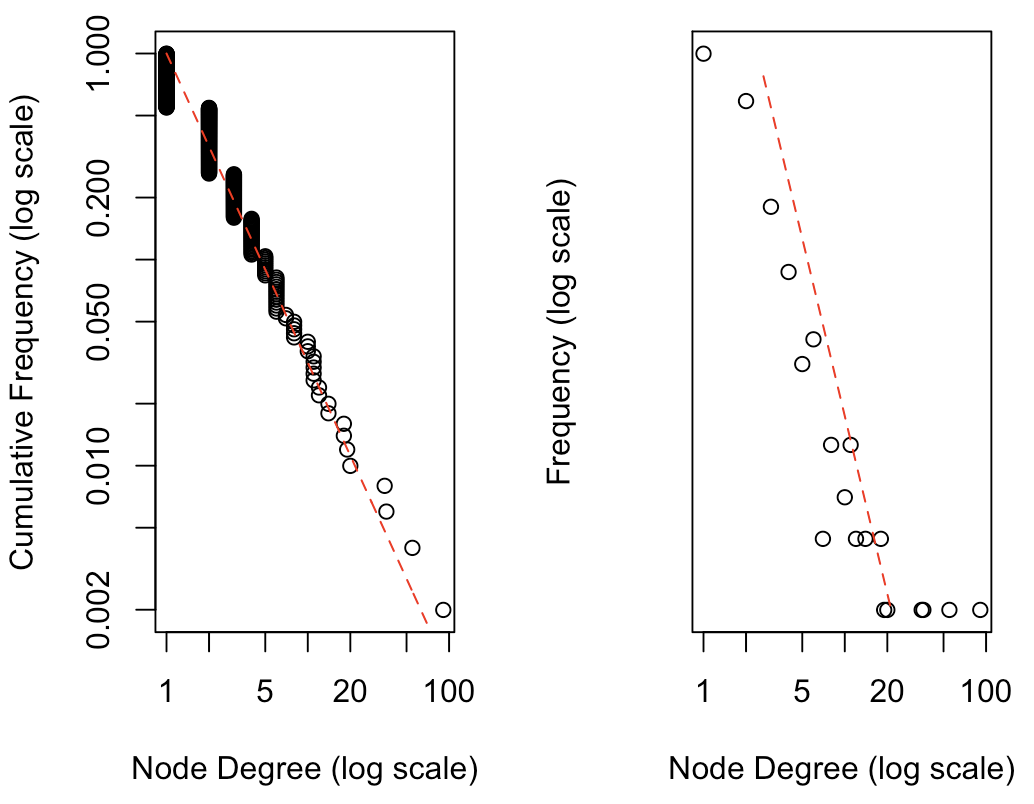}~~    \includegraphics[height=.3\textheight]{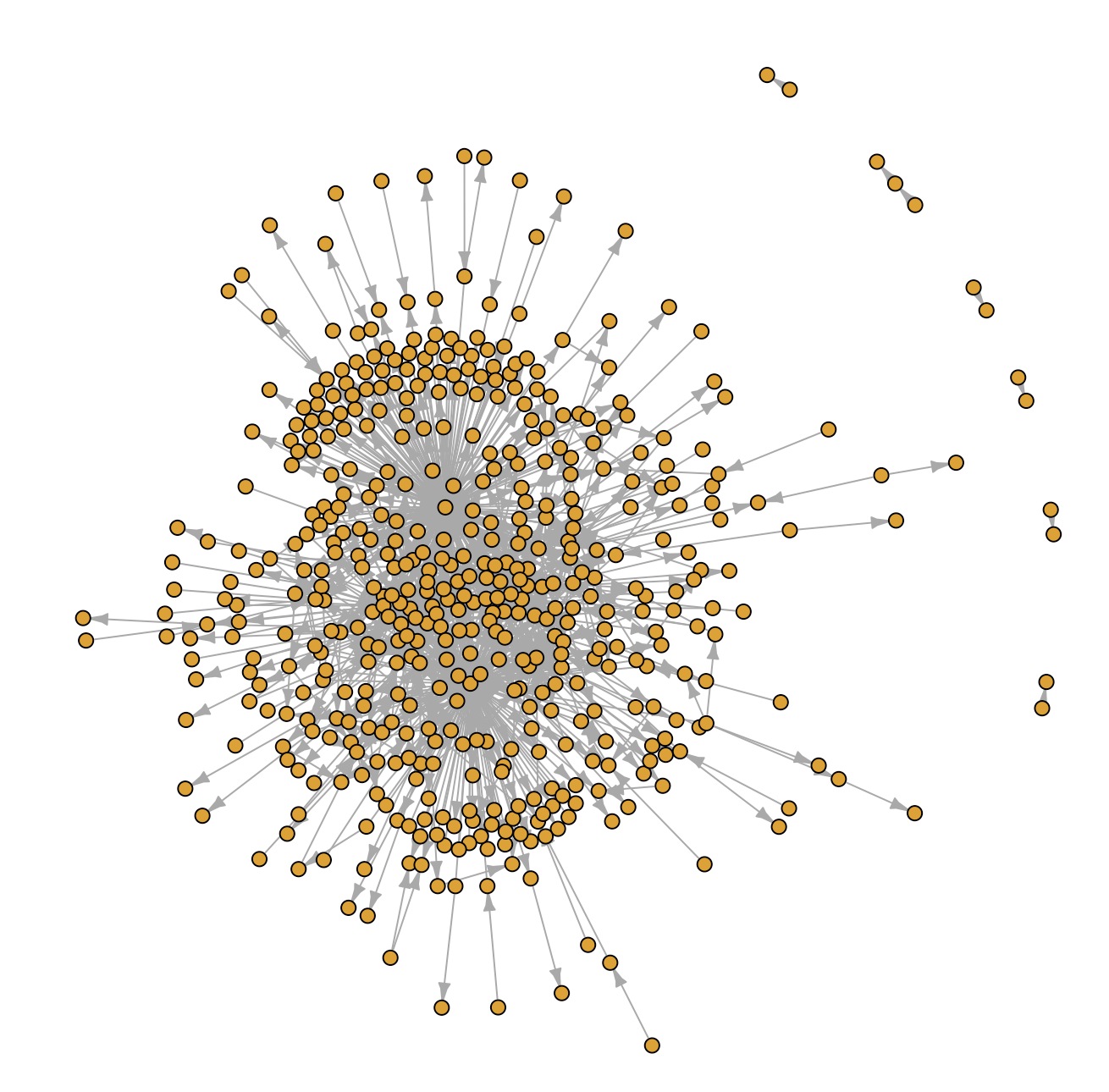}
    \caption{Strict Scale Free Network}
    \label{fig:net1}
\end{figure}

\begin{figure}
    \centering
    \includegraphics[height=.3\textheight]{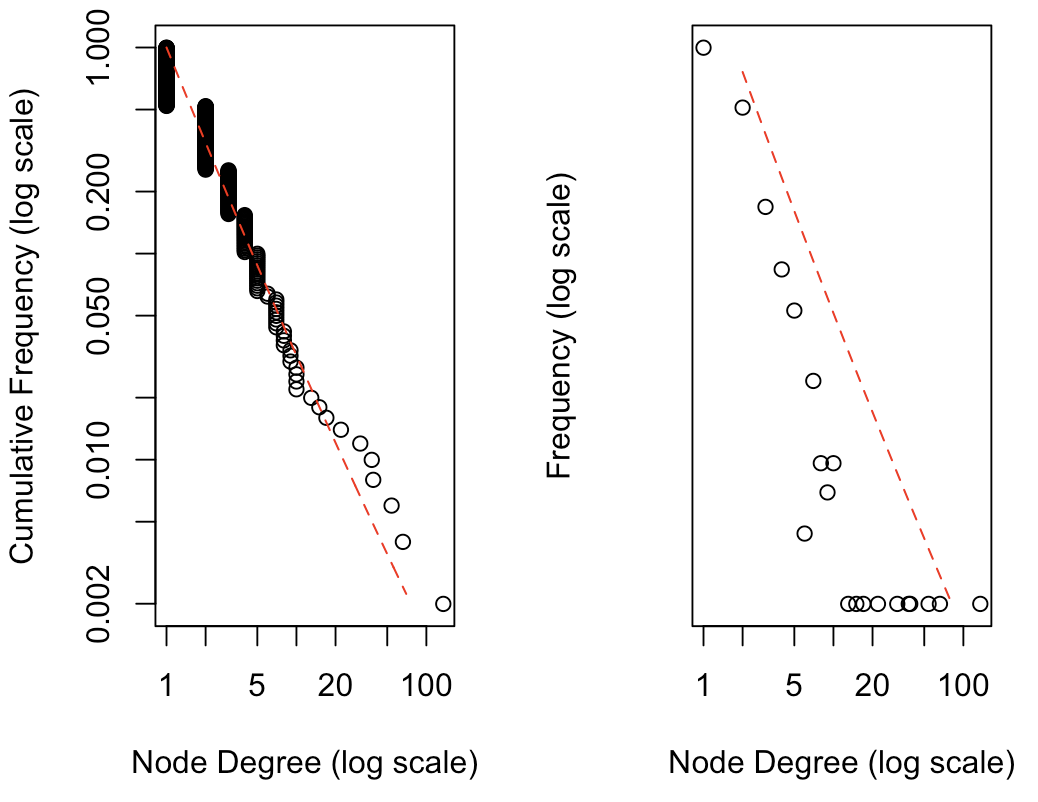}~~    \includegraphics[height=.3\textheight]{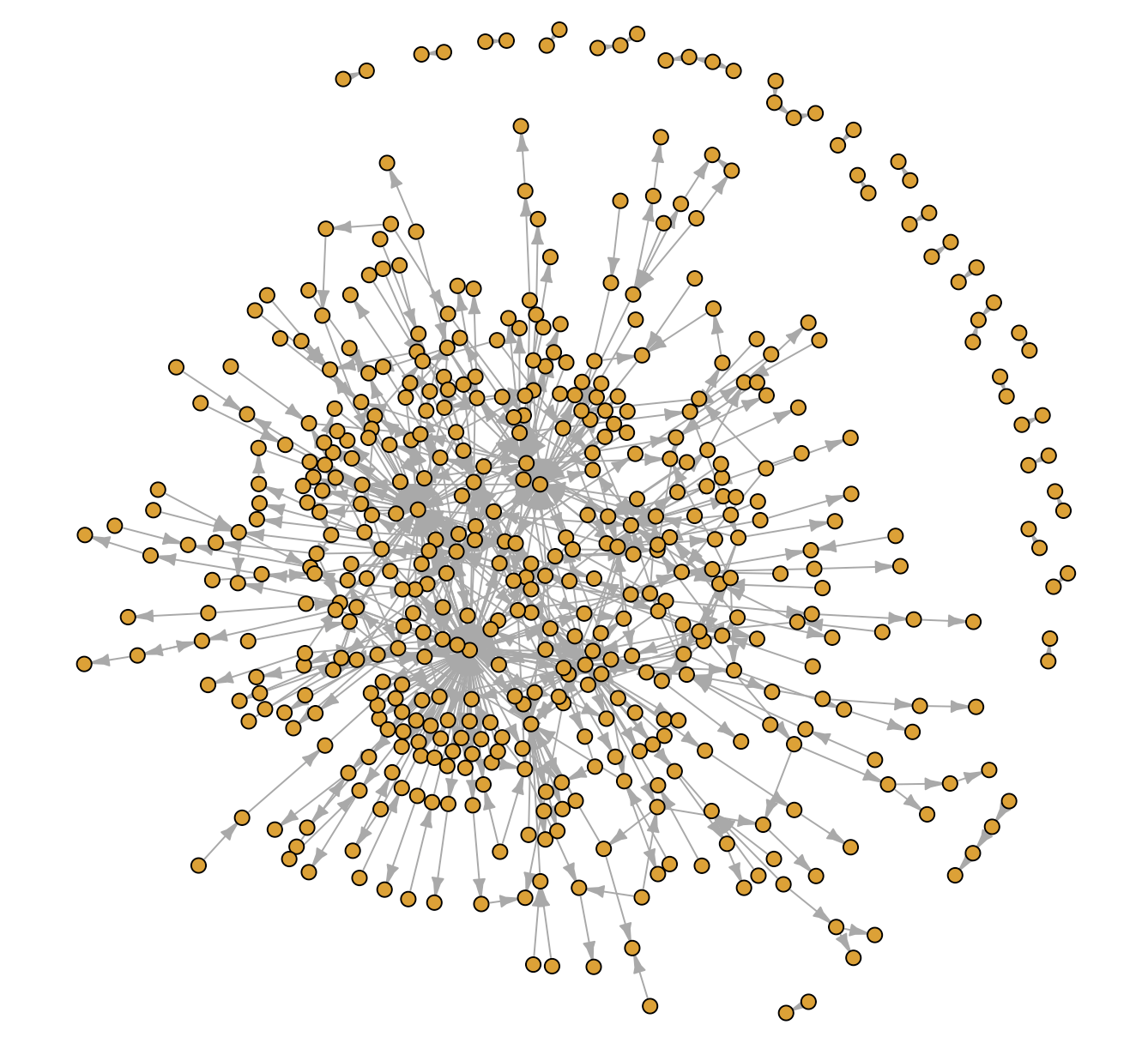}
    \caption{Extended Scale Free Network}
    \label{fig:net2}
\end{figure}

\subsection{Sub-network of Scale-Free Networks}

\cite{Stumpf} claimed that sub-networks of scale-free networks are not scale-free anymore. Of course, this result depend on how we sub-sample from a general network, and how scale-free is defined. Let $\{\mathcal{V},\mathcal{E}\}$ a network, i.e. a collection of vertices and edges. Let $A$ denote its adjacency matrix : let $i$ and $j$ denote two nodes in $\mathcal{V}$, then $A_{i,j}=1$ if an only if $(i-j)$ is in $\mathcal{E}$. We assume here that there are no zero-degree node, i.e. $\forall i\in\mathcal{V}$, $\exists j\in\mathcal{V}$ such that $A_{i,j}=1$.

To generate a sub-network, select randomly (and uniformly) a sub-sample of nodes $\mathcal{V}^*$, then extract the sub-adjacency matrix $A^\star$, and the $(i-j)$ is in $\mathcal{E}$ if an only if $A^\star_{i,j}=1$.
Interestingly, one can easily write the PGF of the degree distribution on the sub-network, 
$$
G^*(s)=G(1-p(1-s))
$$
where $p$ is the probability to keep a given node. Since we excluded orphaned nodes, it is necessary to rescale the PGM, and then
$$
G^*(s)=\frac{G(1-p(1-s))}{1-G(1-p)}
$$

As discussed in \cite{Stumpf}, if we get back to simple scale-free network, we can obtain the following
$$
\mathbb{P}[d=k]\sim\frac{1}{k}, \text{ then }
\mathbb{P}[d>k]\sim\frac{1}{k^2}\text{ while }
\mathbb{P}^*[d^*>k]\sim\frac{1}{k(k-1)}$$
$$
\mathbb{P}[d=k]\sim\frac{1}{k^2}, \text{ then }
\mathbb{P}[d>k]\sim\frac{1}{k^3}\text{ while }
\mathbb{P}^*[d^*>k]\sim\frac{1}{k(k-1)(k-2)}$$
Observe that those two reminds us of  the Hall class (see \cite{Hall}). Hence
$$
\frac{1}{k(k-1)}=[k^2(1-k^{-1})]^{-1}
$$
which is an extended Pareto distribution with index $\tau=-1$.

\section{Real Internet Networks}

In order to illustrate this second order property, we will use data from the Stanford Network Analysis Project (SNAP), from Facebook\footnote{http://snap.stanford.edu/data/ego-Facebook.html}, Twitter\footnote{http://snap.stanford.edu/data/ego-Twitter.html} and Google Plus\footnote{http://snap.stanford.edu/data/ego-Gplus.html} (see \cite{McAuley}). The first one contains 4,039 nodes and 88,234 edges, the second one contains 81,306 nodes and 1,768,149 edges and the third 107,614 nodes and 13,673,453 edges.

\begin{figure}
    \centering
    \includegraphics[height=.16\textheight]{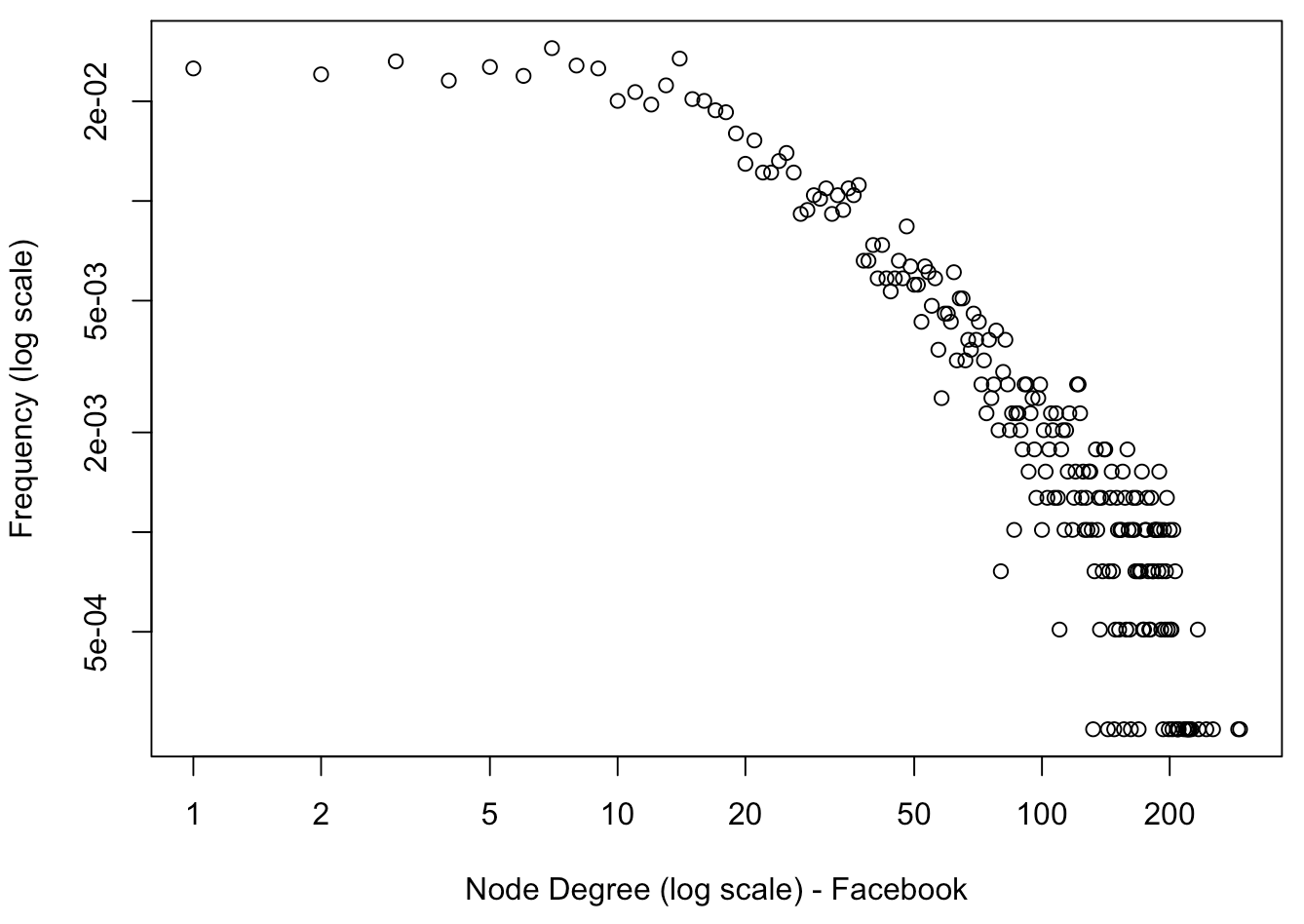}~~   
    \includegraphics[height=.16\textheight]{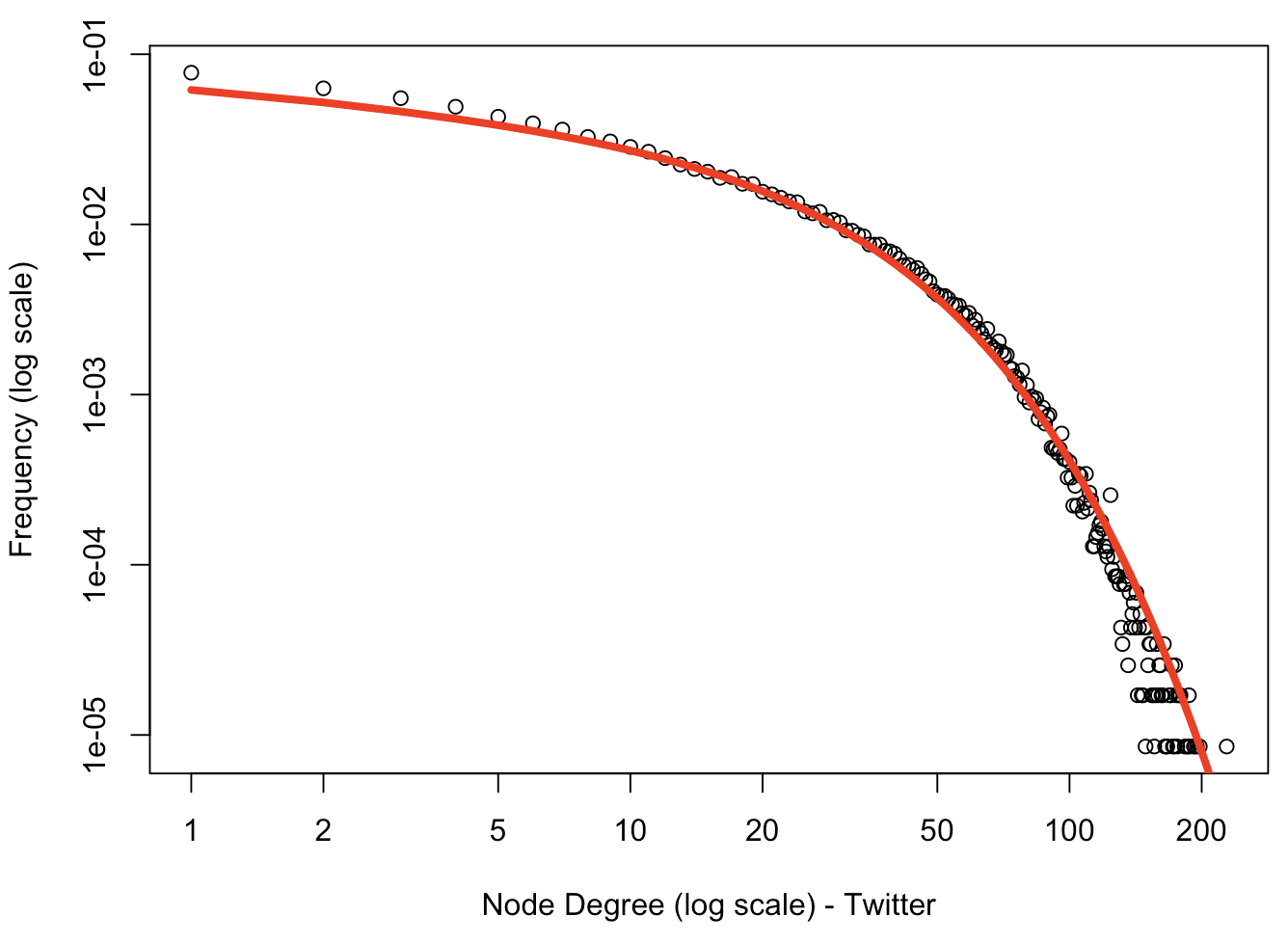}~~   
    \includegraphics[height=.16\textheight]{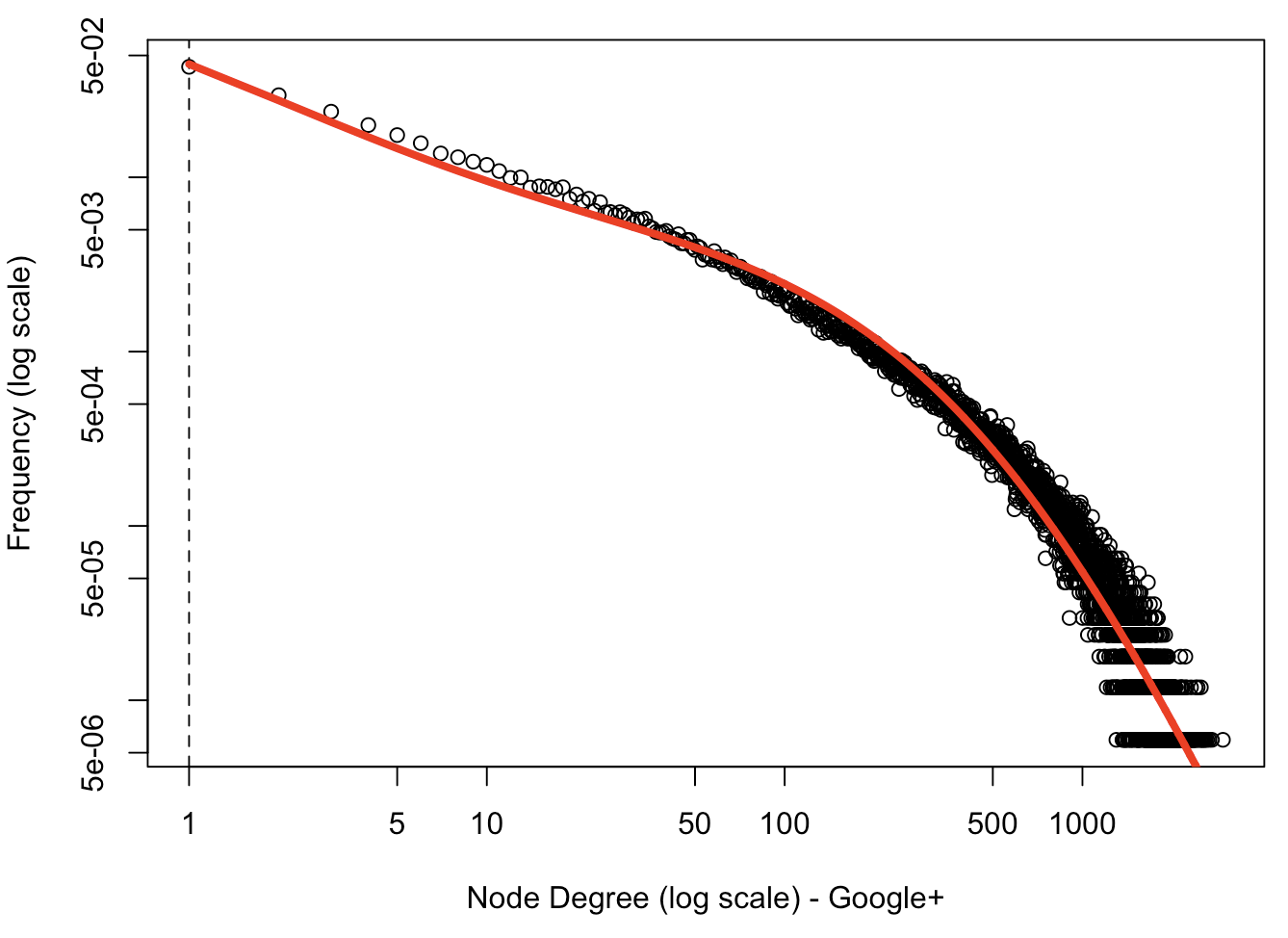} 

    \includegraphics[height=.16\textheight]{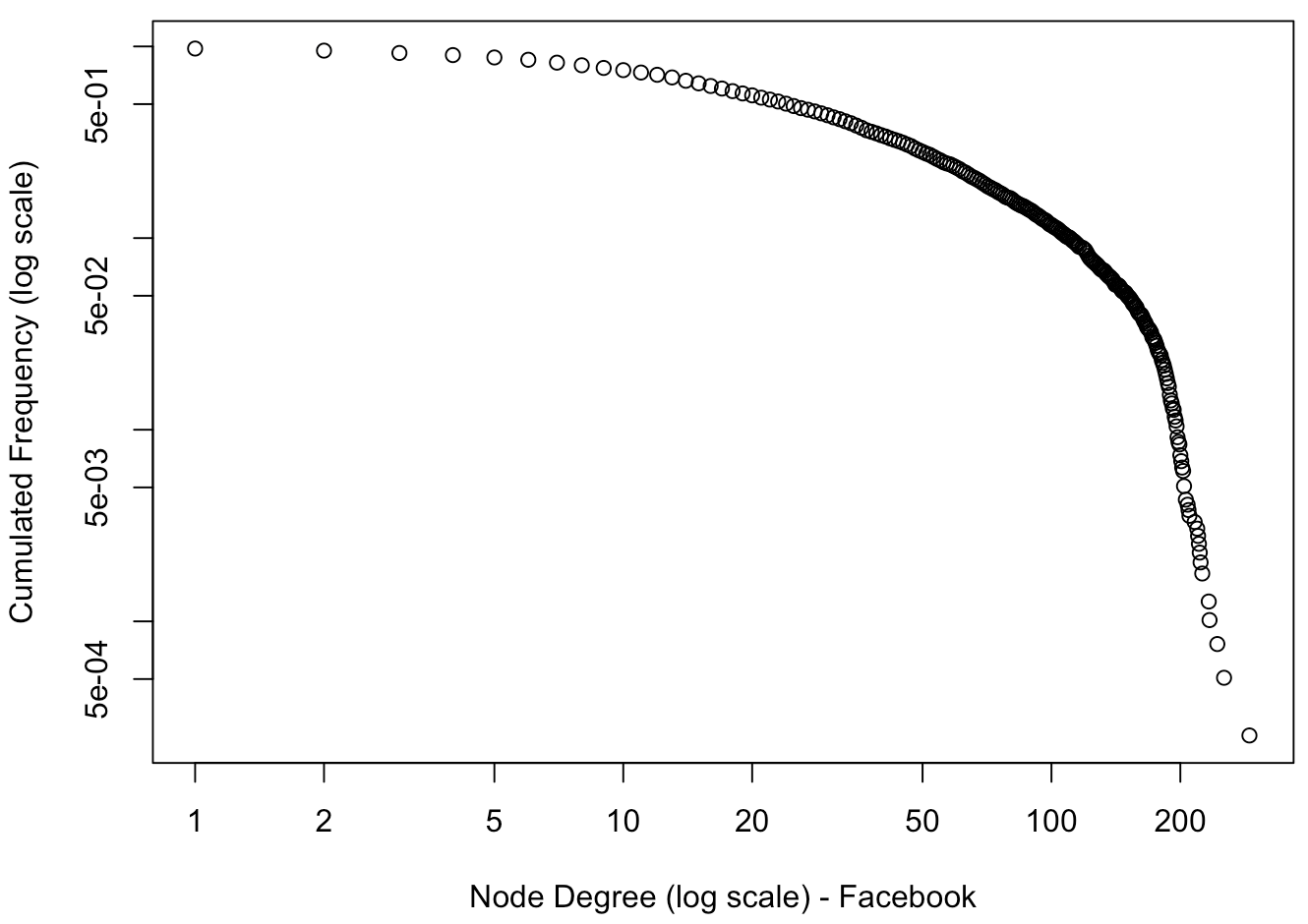}~~   
    \includegraphics[height=.16\textheight]{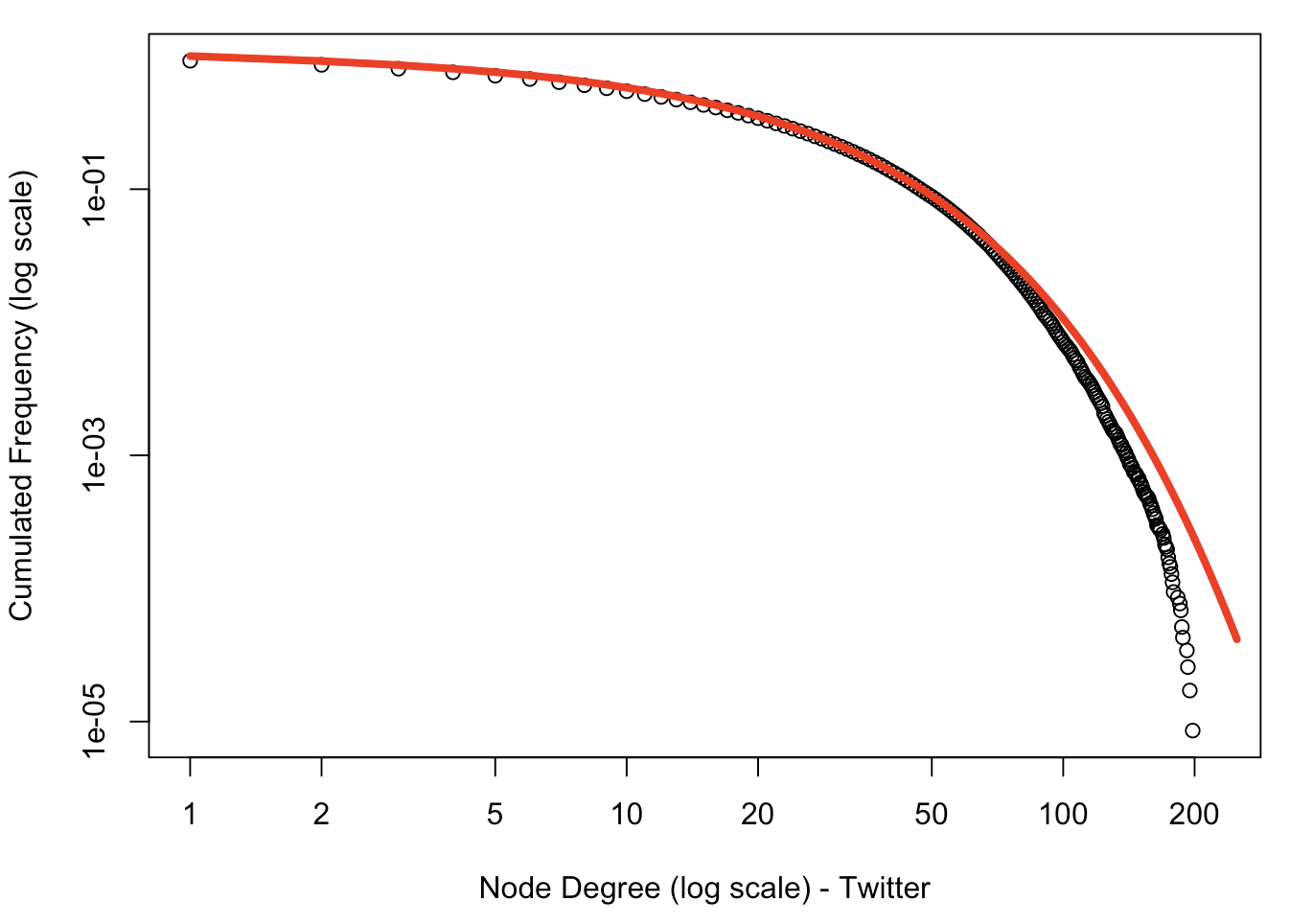}~~   
    \includegraphics[height=.16\textheight]{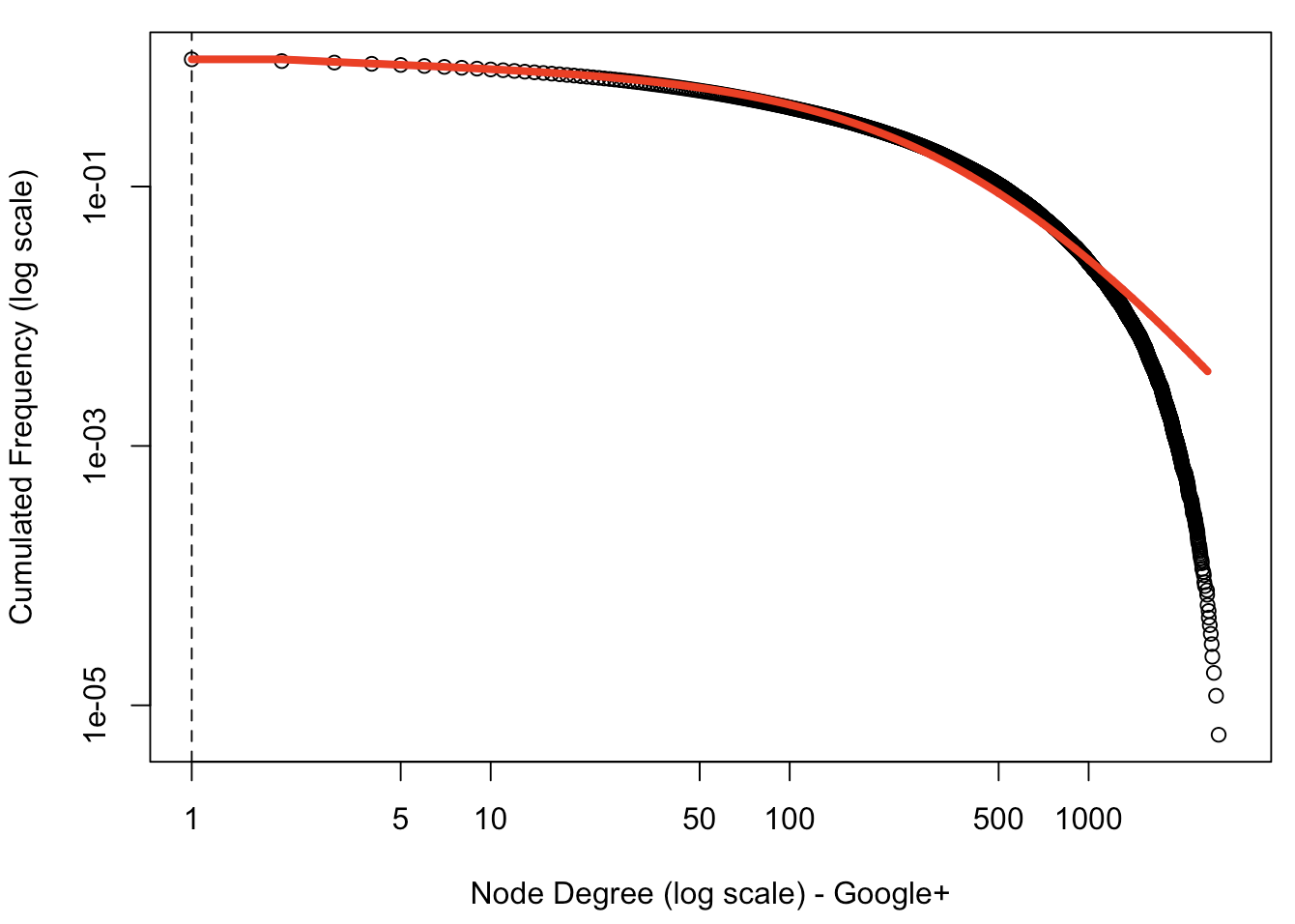} 
    \caption{Distribution of degrees on Facebook sub-network (left), Twitter (center) and Googple+ (right), including the extended scale free distribution (estimated by maximum likelihood) for Twitter and Google+.}
    \label{fig:net2}
\end{figure}

In the SNAP-Facebook dataset, we have we have 10 sub-networks. It is known for being a not scale-free network, see \cite{Gjokaf}. This is confirmed on the left of Figure \ref{fig:net2} where no extended distribution can be used to capture such a strong concavity.

For the Twitter network, \cite{Aparicio} fitted a scale free distribution $\mathbb{P}(k)\sim C k^{-\lambda}$, outgoing degree distribution has tail index of $\widehat{\lambda}=2.1715$ while incoming degree distribution has tail index $\widehat{\lambda}=1.8778$. Here, we did not distinguish incoming and outgoing edges. When fitting an extended Pareto distribution, we obtained $\widehat{\xi}=0.757$ (or  $1+\widehat{\xi}^{-1}=2.32$, consistent with the values obtained in \cite{Aparicio}) and $\widehat{\tau}=-1$. For Google+, we obtained also $\widehat{\tau}=-1$. This value is consistent with the result obtained in \cite{Stumpf}, and can be visualized on the center of Figure \ref{fig:net2} (for Twitter network) and on the right of \ref{fig:net2} (for Google+ network). On those two sets of figures, the parametric fitted distribution is added to the scatterplot.

\section{Appendix}

\subsection{Fitting Continuous Distributions}\label{App1}

Consider a continuous power law distribution, with density
$$
f(x)=(\alpha-1)x^{-\alpha},~x\geq 1.
$$
The likelihood of a sample $\boldsymbol{x}=\{x_1,\ldots,x_n\}$ is then 
$$
\mathcal{L}(\alpha;\boldsymbol{x})=\prod_{i=1}^nf(x_i)=\prod_{i=1}^n(\alpha-1)(x_i)^{-\alpha}
$$
For convenience, use the logarithm of the likelihood,
$$
\log\mathcal{L}(\alpha;\boldsymbol{x})=\sum_{i=1}^n\log(\alpha-1)-\alpha\log(x_i)
$$
The maximum of the logarithm of the likelihood function is obtained when
$$
\left.\frac{\partial \log\mathcal{L}(\alpha;\boldsymbol{x})}{\partial\alpha}\right\vert_{\alpha=\widehat{\alpha}}=0\text{ i.e. }\widehat{\alpha}=1+n\left(\sum_{i=1}^n\log(x_i)\right)^{-1}
$$

\subsection{Fitting Discrete Distributions}\label{App2}

It was mentioned in section \ref{sec:inf:discrete} that the chi-square distance can be used to estimate the (unknown) parameter
$$
Q(\boldsymbol{\theta})=\sum_{k=1}^{k_{\max}}\frac{(np_{d-\star,\boldsymbol{\theta}}(k)-n_k)^2}{np_{d-\star,\boldsymbol{\theta}}(k)}
$$
A more robust version can be obtained by regrouping too-small degrees, to have at least 10 nodes : consider some (consecutive) partition $\mathcal{K}_1,\ldots,\mathcal{K}_m$ such that $\sum_{k\in\mathcal{K}_j}n_k\geq 10$ for all $j$, and then 
$$
Q(\boldsymbol{\theta})=\sum_{j=1}^{m}\frac{(np_{d-\star,\boldsymbol{\theta}}(\mathcal{K}_j)-n_{\mathcal{K}_j})^2}{np_{d-\star,\boldsymbol{\theta}}(\mathcal{k}_j)}
$$ where
$$
n_{\mathcal{K}_j}=\sum_{k\in\mathcal{K}_j}k\text{ and }
p_{d-\star,\boldsymbol{\theta}}(\mathcal{K}_j)=F_{\star}(\max\{\mathcal{K}_j\})-F_{\star}(\min\{\mathcal{K}_j\}-1)
$$
Then the estimator is
$$
\widehat{\boldsymbol{\theta}}=\text{argmin}\{Q(\boldsymbol{\theta})\}
$$

For the discrete EPD model, given a vector \texttt{x} of degrees, the {\sffamily R} code to compute the chi-square distance between the empirical distribution and $p_{d-\text{EPD}}$ is, for some given $\boldsymbol{\theta}$ (i.e. values \texttt{gamma}, \texttt{tau} and \texttt{kappa})
\begin{verbatim}
    T = table(x)
    T2 = T[as.character(1:max(as.numeric(names(T))))]
    names(T2) = as.character(1:max(as.numeric(names(T))))
    T2[is.na(T2)] = 0
    k = 1
    sumt2 = 0
    VK = NULL
    k0 = k
    while(k<=max(as.numeric(names(T)))){
      sumt2=sumt2+T2[as.character(k)]
      if(sumt2<10){k=k+1}
      if(sumt2>=10){VK=rbind(VK,c(k0,k,sumt2))
      k0=k
      k=k+1
      sumt2=0}
    }
    VK[2:nrow(VK),1] = VK[2:nrow(VK),1]+1
    PEMP = VK[,3]/(VK[,2]+1-VK[,1])/sum(VK[,3])
    PEPD = pepd(VK[,2]+1,gamma=gamma,tau=tau,kappa=kappa)-
           pepd(VK[,1],gamma=gamma,tau=tau,kappa=kappa)
    VK = cbind(VK,PEMP,PEPD)
    Q = sum( (PEMP-PEPD)^2/PEPD )
\end{verbatim}
Then we use an optimization route (mainly function \texttt{optim}()) to find $\widehat{\boldsymbol{\theta}}$.

The maximum likelihood is obtained here with a slight change at the end of the previous code
\begin{verbatim}
    PEPD=pepd(x+1,gamma=gm,tau=ta,kappa=kp)-
         pepd(x,gamma=gm,tau=ta,kappa=kp)
    MLE=-sum(log(PEPD))
\end{verbatim}

Then, use \texttt{optim} to find the maximum of the log-likelihood, or the minimum of the chi-square distance.

\begin{figure}
    \centering
    \includegraphics[height=.17\textheight]{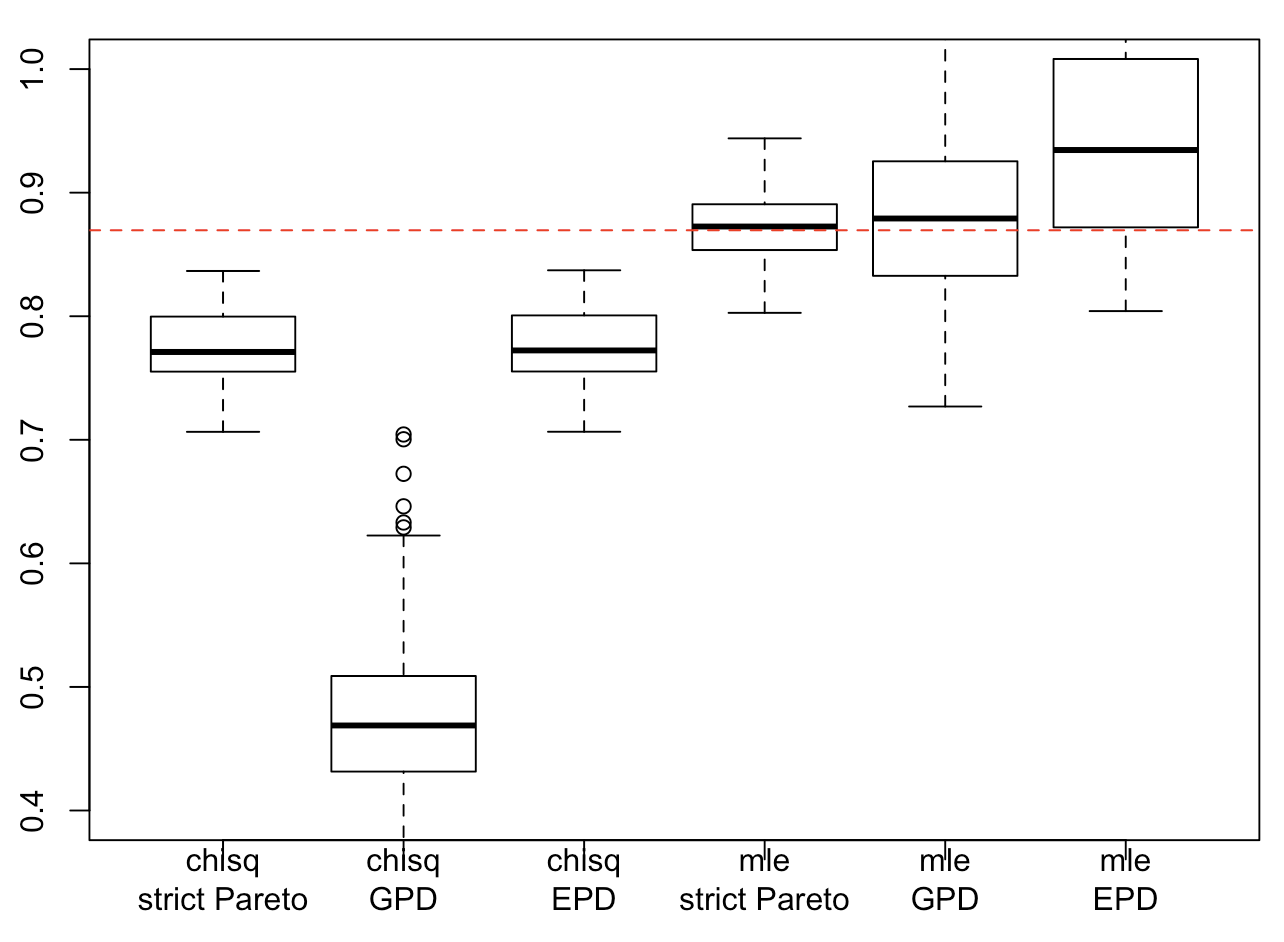}~~    \includegraphics[height=.17\textheight]{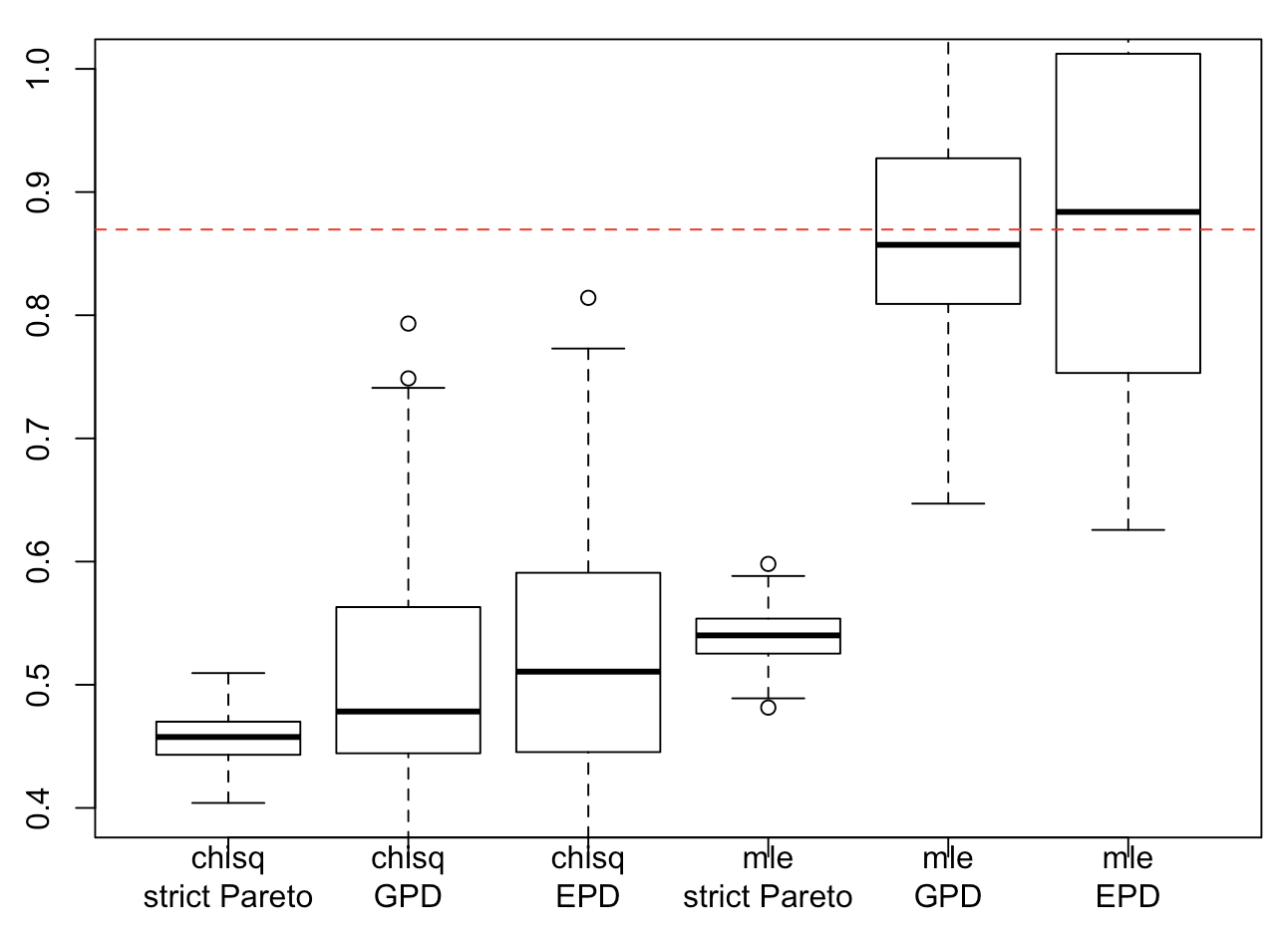}~~    \includegraphics[height=.17\textheight]{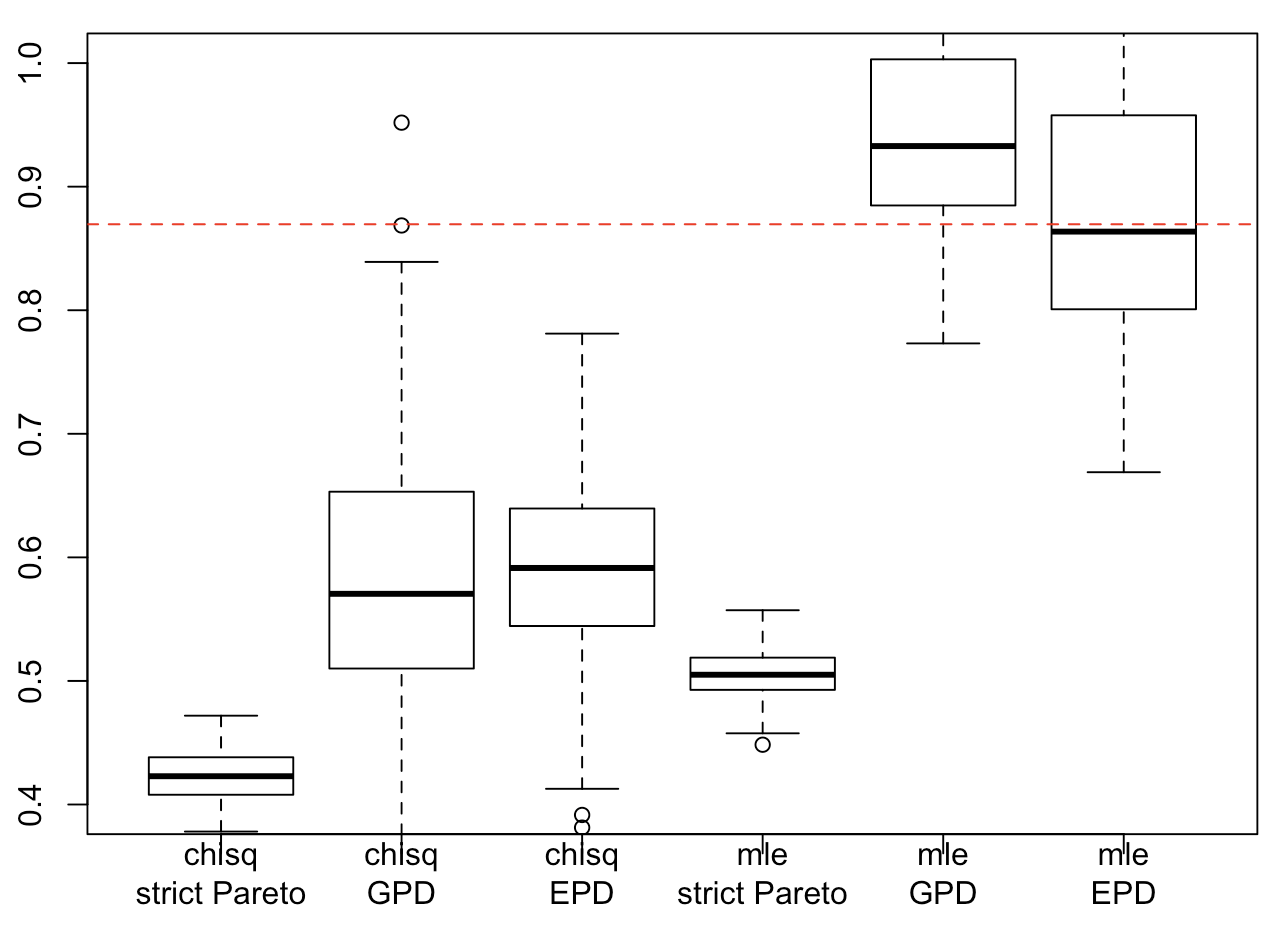}
    \caption{Simulated strict Pareto on the left ($\alpha=1.15$) and extended Pareto in the center ($\alpha=1.15$ and $\tau=-1$) and on the right ($\alpha=1.15$ and $\tau=-1.6$). Boxplot represent the distribution of $\widehat{\xi}$ over $1,000$ simulated networks (with $1,000$ nodes).}
    \label{fig:mle-chi}
\end{figure}

On Figure \ref{fig:mle-chi}, we can visualize the boxplots of the six-estimators of $\alpha$ considered here, on 1,000 simulated samples, with two techniques and three underlying distribution (a discrete strict Pareto with tail index $alpha=1.15$). On the left, we use the chi-square minimum distance, and on the right, the maximum-likelihood technique. We consider either a strict Pareto, a Generalized Pareto (GPD) and the Extended Pareto (EPD).

\end{document}